\documentstyle[epsfig,amssymb,floatflt,float,here]{mn}
\newcommand{\sm}{\, {\rm M}_{\odot}}
\newcommand{\sL}{\, {\rm L}_{\odot}}
\newcommand{\kms}{\,{\rm km \, s^{-1}}}

\newcommand{\ndeg}{^{\circ}}

\title{Simple dynamical models of the Sagittarius dwarf galaxy}
\author[A. Helmi and S.D.M. White]
       {Amina Helmi$^{1,2,3}$ and Simon D.M. White$^{2}$\\
       $^{1}$ Sterrewacht Leiden, Postbus 9513, 2300 RA Leiden, 
The Netherlands\\
	$^{2}$ Max-Planck-Institut f\"ur Astrophysik, Karl-Schwarzschild-Str. 
1, 85740 Garching bei M\"unchen, Germany\\
	${^3}$ Observatorio Astron\'omico de La Plata, Paseo del Bosque s/n,
	B1900FWA La Plata, Argentina}
\date{Accepted ...
      Received ...;
      in original form ...}

\pagerange{\pageref{firstpage}--10}
\pubyear{}

\begin{document}

\maketitle

\label{firstpage}

\begin{abstract}
We present two simple dynamical models for Sagittarius based on N-body
simulations of the progressive disruption of a satellite galaxy
orbiting for 12.5 Gyr within a realistic Galactic potential.  In both
models the satellite initially has observable properties similar to
those of current outlying dwarfs; in one case it is purely stellar
while in the other it is embedded in an extended massive halo.  The
purely stellar progenitor is a King model with a total velocity
dispersion of 18.9 $\kms$, a core radius of 0.44 kpc and a tidal
radius of 3 kpc.  The initial stellar distribution in the other case
follows a King profile with the same core radius, a slightly larger
total velocity dispersion and similar extent. Both these models are
consistent with all published data on the current Sagittarius system, they
match not only the observed properties of the main body of Sagittarius, but
also those reported for unbound debris at larger distances.
\end{abstract}


\section{Introduction}
The Sagittarius dwarf galaxy is the closest satellite of the Milky Way
(Ibata, Gilmore \& Irwin 1994, 1995, hereafter IGI95). Soon after its
discovery, several groups carried out simulations to see if its
properties are consistent with the disruption of an object similar to
the other dwarf companions of the Milky Way, but none produced a model
in full agreement with both the age and the structure of the observed
system (Johnston, Spergel \& Hernquist 1995; Vel\'azquez \& White
1995; Edelsohn \& Elmegreen 1997; Ibata et al. 1997, hereafter I97;
G\'omez-Flechoso, Fux \& Martinet 1999).  All groups assumed light to
trace mass and an initial system similar to observed dwarf
spheroidals. All found the simulated galaxy to disrupt after one or
two orbits whereas the observed system has apparently completed ten or
more. Most considered this to be a problem (but cf Vel\'azquez \&
White 1995). As a result, several unconventional models were proposed
to explain the survival and structure of Sagittarius.  In an extensive
numerical study, Ibata \& Lewis (1998) concluded that Sagittarius must have a
stiff and extended dark matter halo if it is to survive with 25\% of
its initial mass still bound today.  Since an extended halo cannot
remain undistorted in the Galaxy's tidal field for any conventional
form of dark matter, it is unclear how this idea should be
interpreted. Furthermore, it produces an uncomfortably large
mass-to-light ratio ($\sim 100$), it cannot reproduce the observed
elongation, and it suggests that little tidal debris will be
liberated, in apparent conflict with the observations of Mateo,
Olszewski \& Morrison (1998), and Majewski et al. (1999) (see also
Johnston et al. 1999). A somewhat less unorthodox model was proposed
by Zhao (1998), where Sagittarius was scattered onto its current tightly
bound orbit by an encounter with the Magellanic Clouds about 2 Gyr
ago. This appears physically possible but requires careful tuning of
the orbits of the two systems (see Ibata \& Lewis 1998; and Jiang \&
Binney 2000).  Another mechanism by which the dwarf could have moved
to a short-period orbit is dynamical friction, which can be important
only if Sagittarius has lost a lot of mass in the past.  Jiang \&
Binney (2000) found a one-parameter family of initial configurations
that evolve into something like the present system over a Hubble
time. Their initial systems have masses $\sim10^{10-11}\sm$ and start from
a Galactocentric radius $\sim200$ kpc.

Driven by this apparent puzzle, we decided to search more thoroughly
for a self-consistent model of the disruption of Sagittarius, which,
after a Hubble time, has similar characteristics to those
observed. (See Table~\ref{table_sag} for a summary of the observed
properties of the system.) Below we present two models which meet
these requirements.
\begin{table}
\caption{Properties of Sagittarius \, (IGI95, I97)}
\label{table_sag}
\begin{tabular}{lc}
\hline
Orbital properties & \\
\hline
distance from the Sun $d$ & $25 \pm 2$ kpc \\
heliocentric radial velocity $v_r^{\rm sun}$ & $140 \pm 2 \kms$ \\
proper motion in $b$ $\mu_{b}$ & $250 \pm 90 \kms$ \\
gradient along the orbit ${\rm d}v_r/{\rm d}b$ & $< 3 \kms/\deg$ \\
angular position in the sky $(l, b)$ & $(5.6\ndeg,-14\ndeg)$ \\
\hline
Internal properties & \\
\hline
luminosity & $\gtrsim 10^7 \sL$ \\
velocity dispersion $\sigma(v_r)$ & $11.4 \pm 1 \kms$ \\
angular extent in ($l$, $b$) & $8 \ndeg \times 3 \ndeg$ \\
half-mass radius & 0.55 kpc \\
mean metallicity $\langle$[Fe/H]$\rangle$ & $\sim -1.$ dex \\
\hline
\end{tabular}
\end{table}

\section{Method}

In our numerical simulations, we represent the Galaxy by a fixed potential
with three components: a dark logarithmic halo
\begin{equation}
\label{sag_eq:halo}
\Phi_{\rm halo} = v^2_{\rm halo} \ln (r^2 + d^2),
\end{equation}
a Miyamoto-Nagai disk
\begin{equation}
\label{sag_eq:disk}
\Phi_{\rm disk} = - \frac{G M_{\rm disk}}{\sqrt{R^2 + (a + 
\sqrt{z^2 + b^2})^2}},
\end{equation}
and a spherical Hernquist bulge
\begin{equation}
\label{sag_eq:bulge}
\Phi_{\rm bulge} = - \frac{G M_{\rm bulge}}{r + c}, 
\end{equation}
where $d$=12 kpc and $v_{\rm halo}$~=~131.5 $\kms $; $M_{\rm disk} =
10^{11} \sm $, $a$ = 6.5 kpc and $b$~=~0.26 kpc; $M_{\rm bulge} = 3.4
\times 10^{10} \sm$ and $c$~=~0.7 kpc. This choice of parameters gives
a flat rotation curve with an asymptotic circular velocity of 186
$\kms$. The mass of the dark-matter halo within 16 kpc is 7.87 $\times
10^{10} \sm$ in this model. 

We represent the satellite galaxy by a collection of $10^5$ particles
and model their self-gravity by a multipole expansion of the internal
potential to fourth order \cite{W83,Zaritsky_White}.  This type of
code has the advantage that a large number of particles can be
followed in a relatively small amount of computer time. Hence a
substantial parameter space can be explored while retaining
considerable detail on the structure of the disrupted system.  In this
quadrupole expansion, higher than monopole terms are softened more
strongly. We choose $\epsilon_1 \sim 0.2 - 0.25 r_c$ for the monopole
term ($r_c$ is the core radius of the system) and $\epsilon_2 = 2
\epsilon_1$ for dipole and higher terms and for the centre of
expansion. The centre of expansion is a particle which, in practice,
follows the density maximum of the satellite closely at all times.

For the stellar distribution of the pre-disruption dwarf we choose a
King model (King 1966), since this is a good representation of the
distant dwarf spheroidals.  King models are defined by a combination
of three parameters: $\Psi(r\!=\!0)$ (depth of the potential well
of the system), $\sigma^2$ (measure of the central velocity
dispersion), and $\rho_0$ (central density) or $r_0$ (King radius).
The ratio $\Psi(r\!=\!0)/\sigma^2$ defines how centrally concentrated
the system is, and for any value of this parameter, a set of
homologous models with different central densities and core (or King)
radii may be found. We assume that the progenitor of Sagittarius obeys the
known metallicity-luminosity relation for the Local Group dSph
\cite{Mateo}.  The metallicity determinations for Sagittarius \cite{IWGIS}
indicate $\langle [{\rm Fe/H}]\rangle  \sim -1$, corresponding to a {\it total}
luminosity in the range $3.5 \times 10^7 - 3.5 \times 10^8 \sL$.  To
obtain an initial guess for the mass of the system, we transform this
luminosity into a mass assuming a mass-to-light ratio $\sim 2$. The
relevant initial stellar mass interval is then $7 \times 10^7 - 7
\times 10^8 \sm$.

Note that our choice of a fixed potential to represent our Galaxy means
that we neglect any exchange of energy between the satellite and
the Galactic halo. This is an excellent approximation for the range of
orbits and satellite masses that we consider, since these imply
dynamical friction decay times substantially in excess of the Hubble
time. The orbits are also sufficiently large that impulsive heating
during disk passages can be neglected.

The orbit of Sagittarius is relatively well constrained \cite{IWGIS}.
The heliocentric distance $d \sim 25 \pm 2$ kpc and position $(l,b) =
(5.6\ndeg, -14\ndeg)$ of the galaxy core are well-determined; the
heliocentric radial velocity $v_r^{\rm sun} \sim 140 \pm 2\kms $, and
its variation across the satellite are also accurately
measured. Outside the main body ($b < - 20\ndeg$) the radial velocity
shows a small gradient ${\rm d}v_r/{\rm d}b \lesssim 3 \kms \deg$, but
no gradient is detected across the main body itself. The proper motion
measurements are not very accurate; $\mu_b \sim 2.1 \pm 0.7 {\rm
mas\,yr^{-1}}$, and no measurement is available in the
$l$-direction. On the other hand the strong North-South elongation of
the system suggests that it has little motion in the $l$-direction,
thus implying the orbit should be close to polar. We generate a
range of possible orbits satisfying these constraints and concentrate
on those with relatively long periods in order to maximise the
survival chances of our satellite. We begin all our simulations half a
radial period after the Big Bang to allow for the initial
expansion. We place the initial satellite at apocentre, then we
integrate forward until $\sim 13$ Gyr. The orbits are chosen so at
this time the position and velocity of the satellite core correspond
to those observed. We allow ourselves some slight freedom in choosing
the final time in order to fit the observed data as well as possible.

\begin{figure}
\begin{center}
\psfig{figure=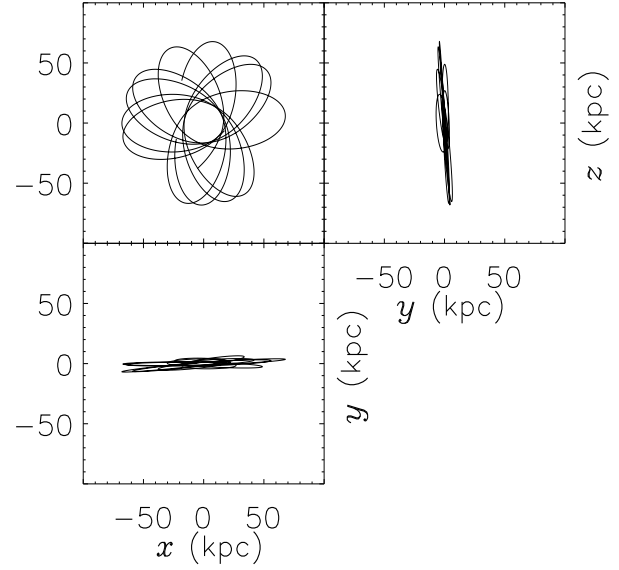,height=8cm,width=8.cm}
\caption{Projections of a possible orbit of Sagittarius on different 
orthogonal planes, where $xy$ coincides with the plane of the Galaxy.}
\label{fig:orbit_sag}
\end{center}
\end{figure}

\section{Results}

Figure~\ref{fig:orbit_sag} gives an example of an orbit which is
consistent with all the current data on Sagittarius.  It has a pericentre of
$16.3$ kpc, an apocentre of $68.3$ kpc, and a radial period of $\sim
0.85$ Gyr. We use similar orbits for all the simulations described below.
Note that the slow precession about the Galactic rotation axis is in part
due to the quasi-polar nature of the orbit and in part to the fact we have
assumed the Milky Way's dark halo to be spherical. 

After letting our satellite relax in isolation, we integrate each
simulation for $\sim 13$ Gyr. In practice we needed to run a large
number of simulations, and test each to see if it satisfies the
observational constraints at the present time. Since it remains
uncertain whether dwarf spheroidals have extended dark halos
(e.g. Klessen \& Kroupa 1998), we have considered both purely stellar
models and models in which the initial stellar system is embedded in a
more massive and more extended dark halo.

\subsection{Constant mass-to-light ratio: A purely stellar model}

Our preferred purely stellar model (Model I) initially has a core
radius of $r_c$ = 0.44 kpc, a total velocity dispersion of 18.9
$\kms$, and a concentration parameter $c = \log_{10}(r_t/r_c) \sim
0.83$. This implies a total mass  of $ M = 4.66 \times 10^8 \sm$.
For a satellite to survive for about 10 Gyr on an orbit with pericentre
$\sim 15$ kpc, apocentre $\sim 70$ kpc, and period $\sim 1$ Gyr (for
which the observational constraints are satisfied) its initial central
density has to be $\rho_0 \geq 0.36 - 0.4 \sm {\rm pc^{-3}}$.
Satellites with significantly smaller initial densities do not survive
long enough.

\begin{figure}
\begin{center}
\psfig{figure=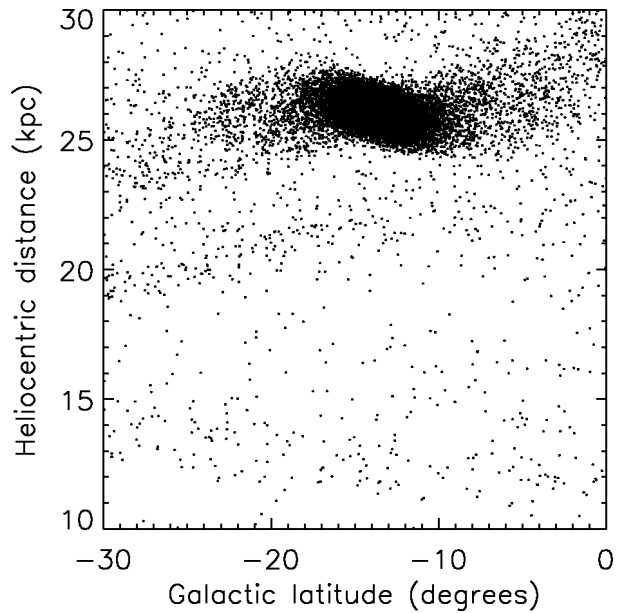,height=8.cm,width=8.cm}
\caption{Distribution of particles in distance from the Sun as a function of
latitude. For direct comparison see Figure~4 of I97.}
\label{fig:lat_dist_sag}
\end{center}
\end{figure}

In Figure~\ref{fig:lat_dist_sag} we plot heliocentric distance as
a function of galactic latitude for stars projected near the main remnant
12.5 Gyr after infall. Streams of particles are visible at all 
latitudes over a broad range in distance. Sagittarius has been orbiting long 
enough for its debris streams to be wrapped several times around the
Galaxy. (See also Figure \ref{fig:all_lat_long_dist_radvel_sgr}.)

\begin{figure*}
\begin{center}
\psfig{figure=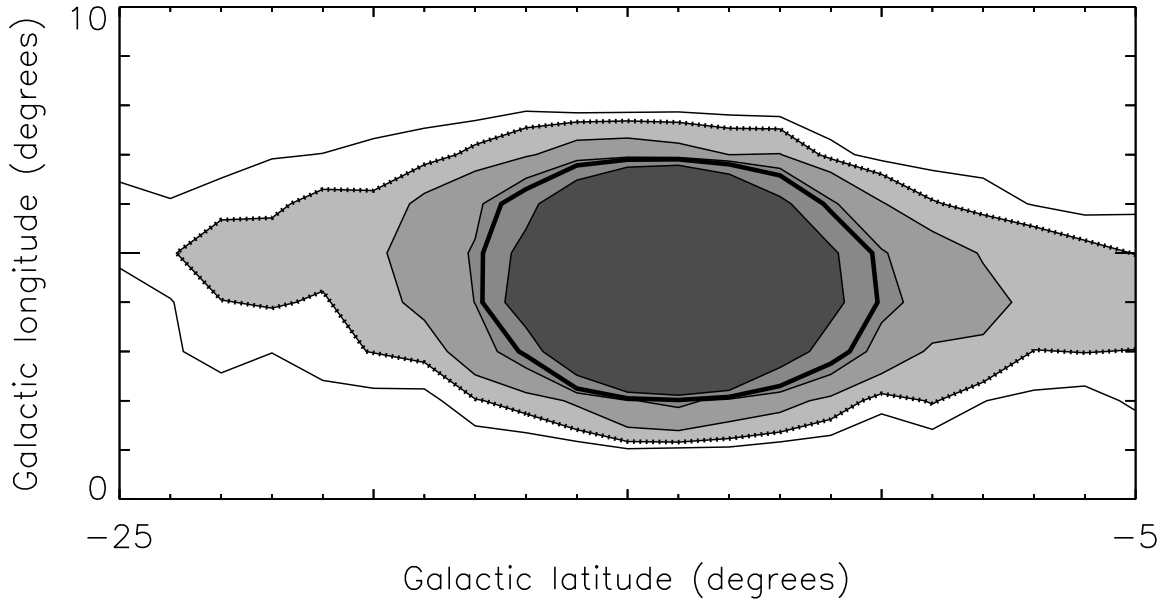,height=17cm,angle=90}
\caption{Surface isodensity contours for the remnant system. The thick
and dashed lines indicate the contours that, for $M/L = 2.25$, would
correspond to the minimum contours plotted in 1994 and in 1997
respectively by Ibata and collaborators. Each succeeding contour has
half the mass surface density of the previous one.}
\label{fig:lat_long_sag}
\end{center}
\end{figure*}

The remnant galaxy, i.e. the central region of the satellite's debris,
is similar to the real system.  In Figure~\ref{fig:lat_long_sag} we
plot its mass surface density.  The transformation from observed
surface brightness to mass surface density (which is what the
simulations give us) can be done as follows.  The observed mass
surface density $\Sigma$ for an assumed mass-to-light ratio
$\Upsilon$~is
\begin{equation}
\Sigma = \frac{N_X L_X }{f_X}\Upsilon \qquad \left[\frac{\sm}{\deg^2}\right],
\end{equation}
where $N_X$ is the number of observed stars of type $X$ per square
degree, $L_X$ is their luminosity, and $f_X$ is the fraction of the
total luminosity in stars of type $X$. In IGI95 the spatial structure
of Sagittarius was determined from the excess of counts at the apparent
magnitude of the horizontal branch. Uncertainties in the result are
due primarily to contamination by sources in the Galactic bulge.
Their lowest isodensity contour is at $\Sigma_{\rm min} \sim 5 \times
10^5 \frac{\sm}{\deg^2}$, assuming $\Upsilon\sim 2.25$ and [Fe/H]
$\sim -1$ (Bergbusch \& vandenBerg 1992), and has an extent of
$7.5\ndeg \times 3\ndeg$. This same isodensity contour is shown in
Figure \ref{fig:lat_long_sag} as a thick line. It has an extent of
$\sim 8\ndeg \times 4.8\ndeg$, in reasonable agreement with the
observations given the uncertainties. In I97 isodensity contours were
derived from counts of main sequence stars close to the turn-off,
roughly one magnitude above the plate limit. The minimum contour in
this case corresponds to $\Sigma_{\rm min} \sim 10^5
\frac{\sm}{\deg^2}$, and has an extent of roughly $15 \ndeg \times
7\ndeg$. In Fig.~\ref{fig:lat_long_sag} this contour is shown as a
dashed-line, and has an extent of $21 \ndeg \times 6.5 \ndeg$, also in
good agreement with the observations.
Note that the isophotes (or isodensity contours) become rounder
towards the centre of the satellite. Its angular core radius is $R_{c}
\sim 1.29\ndeg$, which for a distance of 26 kpc (derived from the
simulations) corresponds to 0.58 kpc, again in good agreement with the
observations.

\begin{figure*}
\begin{center}\vspace*{-1.25cm}
\psfig{figure=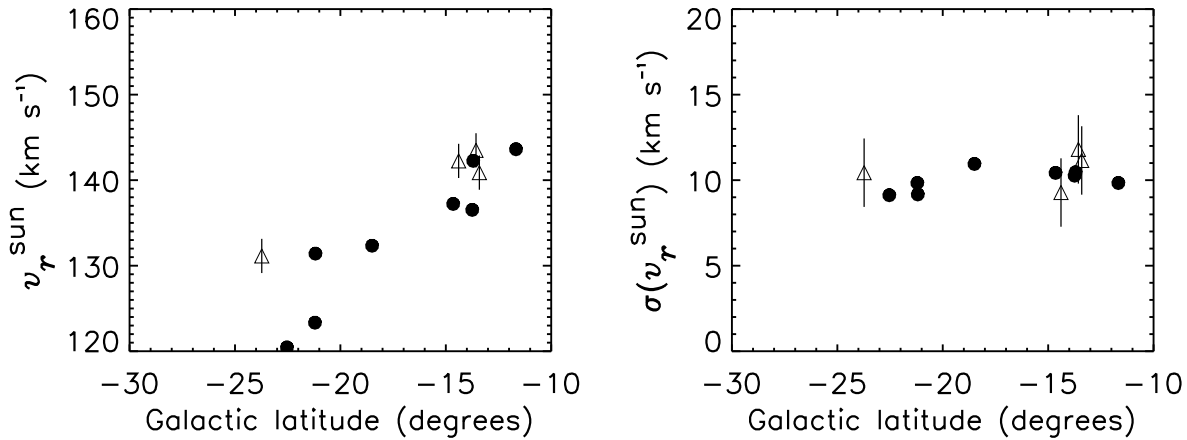,width=16cm}
\caption{In the left panel we plot mean heliocentric radial
velocity as a function of Galactic latitude, for bins of $\sim 2.5\ndeg
\times 2.5\ndeg$ across the remnant system. The right panel shows the
heliocentric radial velocity dispersion in the same bins.  To
determine variations across the main body of the galaxy, we have taken
bins centered on the same Galactic latitude but offset in Galactic
longitude. The triangles correspond to data from I97, error bars
indicate 2 $\kms$ uncertainty.}\label{fig:lat_radvel_sag}
\end{center}
\end{figure*}

The kinematic properties of the remnant galaxy are more difficult to
compare with observations because a substantial amount of mass from
debris streams is projected on top of the main body.  Like I97, we
measure the radial velocity across the system considering only
particles for which $100 \kms \le v_r^{\rm sun} \le 180 \kms$.  In the
left panel of Figure~\ref{fig:lat_radvel_sag} we plot the heliocentric
radial velocity, and in the right panel we plot its dispersion as a
function of Galactic latitude.  For comparison, we analysed the
observations of I97 at CTIO in the same way (their Table 2b); these
data have a precision of a few $\kms$ (triangles in
Figure~\ref{fig:lat_radvel_sag}).  Our model is consistent with the
observed kinematics; we obtain a heliocentric radial velocity of $139.5
\kms$ and an internal velocity dispersion in the radial direction of
$11 \kms$ for the main body. However, when the radial velocity
restrictions for inclusion in this calculation are relaxed, we find
much larger velocity dispersions because of the contribution of stars
from other streams.  It is important to consider this problem when
determining which stars should be considered members of Sagittarius.

\subsection{Varying mass-to-light ratio: A model with a dark halo}

The observational data for Sagittarius mainly refer to the current remnant
system, which corresponds to the innermost regions of the progenitor
satellite. As a consequence, models that are initially dark matter
dominated in their outskirts are relatively poorly constrained.

As an example we focus on a progenitor with a mass distribution which
is similar to that of Model I in its inner regions, but is
considerably more extended.  We take the mass distribution to be a
(heavy) King model with $r_c=0.54$~kpc and $r_t=10.4$~kpc, with an
initial total velocity dispersion of 25.2~$\kms$, and total mass of $
M = 1.7 \times 10^9 \sm$.  For an orbit like that of Model I this
produces a suitable remnant after 12 Gyr.  The {\it mass} distribution
of this remnant satisfies many of the observational constraints of
Table \ref{table_sag}. Its core radius is slightly larger $r_c \sim
0.65$ kpc, and the radial velocity dispersion in the main body is 12.1
$\kms$.

\begin{figure}\hspace*{-0.5cm}
\psfig{figure=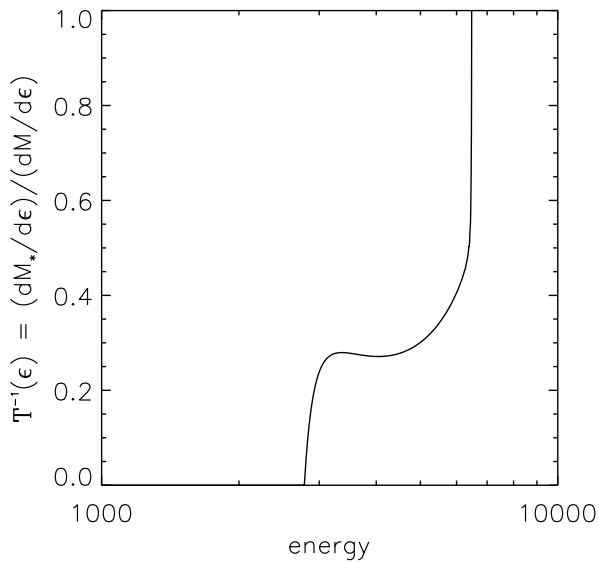,height=8.25cm}
\caption{Inverse mass-to-light ratio as a function of binding energy
for our model II.  Negative values of the energy correspond to unbound
material. Particles in the deepest parts of the potential well have
stellar mass-to-light ratios.}\label{fig:gamma_sag}
\end{figure}

We will construct a two-component satellite with this mass
distribution by solving for the dependence of mass-to-light ratio on
initial binding energy that produces the initial light profile of
Model I. We choose the mass-to-light ratio of satellite material to be
a decreasing function of binding energy, so that the most bound
particles have near ``stellar'' mass-to-light ratios, whereas weakly
bound particles are almost entirely ``dark''.  From the energy
distribution of the heavy King model, and that of a King model with
$r_0 = 0.095$ kpc and $\sigma = 25.6 \kms$, we can derive the
mass-to-light ratio as a function of binding energy as
\begin{equation}
\label{eq:inverse_mass-to-light}
\Upsilon(\epsilon) = \Upsilon_*\frac{{\rm d}M/{\rm d}\epsilon (\epsilon)}
{{\rm d}M_*/{\rm d}\epsilon (\epsilon = \epsilon_* + \epsilon_{\rm max} -
\epsilon_{* \rm max})}
\end{equation}
where $\Upsilon_*$ is the mass-to-light ratio of a stellar population. 
The energies $\epsilon_*$ of the lighter King model have been shifted  
by a fixed amount $\epsilon_{\rm max} - \epsilon_{* \rm max}$, to be
on the same scale as that of the heavier King model. 
The resulting mass-to-light ratio is shown in Figure~\ref{fig:gamma_sag}.

\begin{figure} \hspace*{-0.5cm}
\psfig{figure=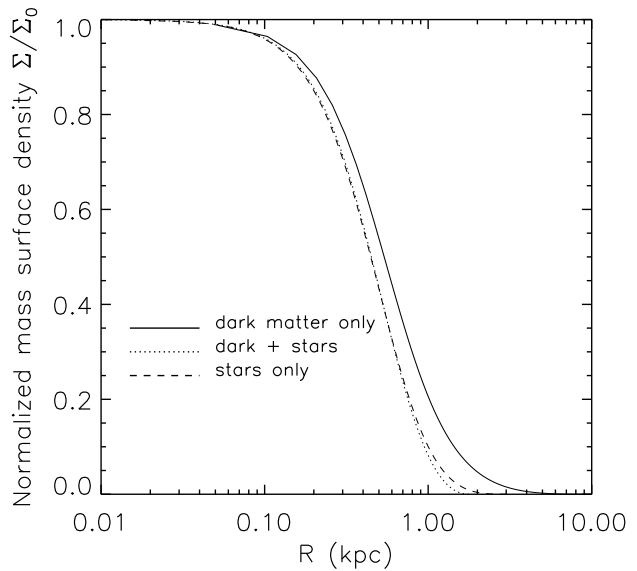,height=8.25cm}
\caption{The dashed curve corresponds to the surface density of Model
I normalized to its central value ($\Sigma_0 = 7.29 \times 10^2
\sm/{\rm pc}^2$).  The solid curve to that of the heavy King model,
which corresponds to the total mass of Model II ($\Sigma_0 = 11.17
\times 10^2 \sm/{\rm pc}^2$). Model II, obtained by weighting each
simulation particle by $(M/L)^{-1}$, is shown as the dotted
curve. Model I and Model II have almost the same
surface density profile by construction.}\label{fig:surf_plot_sag}
\end{figure}

In Figure~\ref{fig:surf_plot_sag} we show the surface mass densities
normalized to their central values for Model~I (only stars), for the
heavy King model and for the two-component model (``stars'' and
dark-matter).  We shall refer to this two-component model as Model~II,
which is obtained by weighting each simulation particle by
$\Upsilon(\epsilon)^{-1}$.

If we require that the central stellar mass surface densities of
Model~I and Model~II be the same, we find that the total mass in stars
in Model~II is $\sim 1.69 \times 10^8 \sm$.  To match Sagittarius  surface
brightness, we choose the central stellar mass-to-light ratio
$\Upsilon_* = 1.5$. Thus, the total luminosity of Model~II is
then $1.13 \times 10^8 \sL$, implying a mass-to-light ratio of
15.1. Its initial velocity dispersion is 23 $\kms$. The visible extent
of the remnant has properties which are almost identical to those of
Model~I, and we find its velocity dispersion to be $11.1 \kms$. Both
results are again in good agreement with the observations.

The two initial satellites (Models~I and II) have the same stellar
mass distributions in their inner regions, differing only in that one
has an extended dark halo. We may thus conclude that the presence of a
dark halo does not affect the final structure of the remnant, which is
very similar in both models. However there is a significant difference
in the properties of their debris streams. In Model I the unbound
debris streams are predicted to contain 5.2 times the light in the
main body of the remnant ($M_V \sim -14.1$), as defined by the dotted
contour in Figure~\ref{fig:lat_long_sag}, whereas in Model II ($M_V
\sim -13.4$) this ratio is 4.85. If we had chosen Model~II to be a
constant mass-to-light ratio model, we would have got an almost
equally good fit to the main body of Sagittarius, but would have predicted
the streams to contain 19 times the light in the main body of the
remnant. In this last case, Sagittarius would have contributed $4.56 \times
10^8 \sL$ to the Galactic stellar halo in the form of debris stars
(for $\Upsilon = 3.5$). Thus we see that the observed properties of
the main remnant do not usefully constrain the number of stars that
may be present in the debris streams, but that the different models
can be better constrained from the properties of their debris streams,
as we exemplify below.

\subsection{Discussion} 
\subsubsection{Some predictions} 

\begin{figure*}
\begin{center}
\psfig{figure=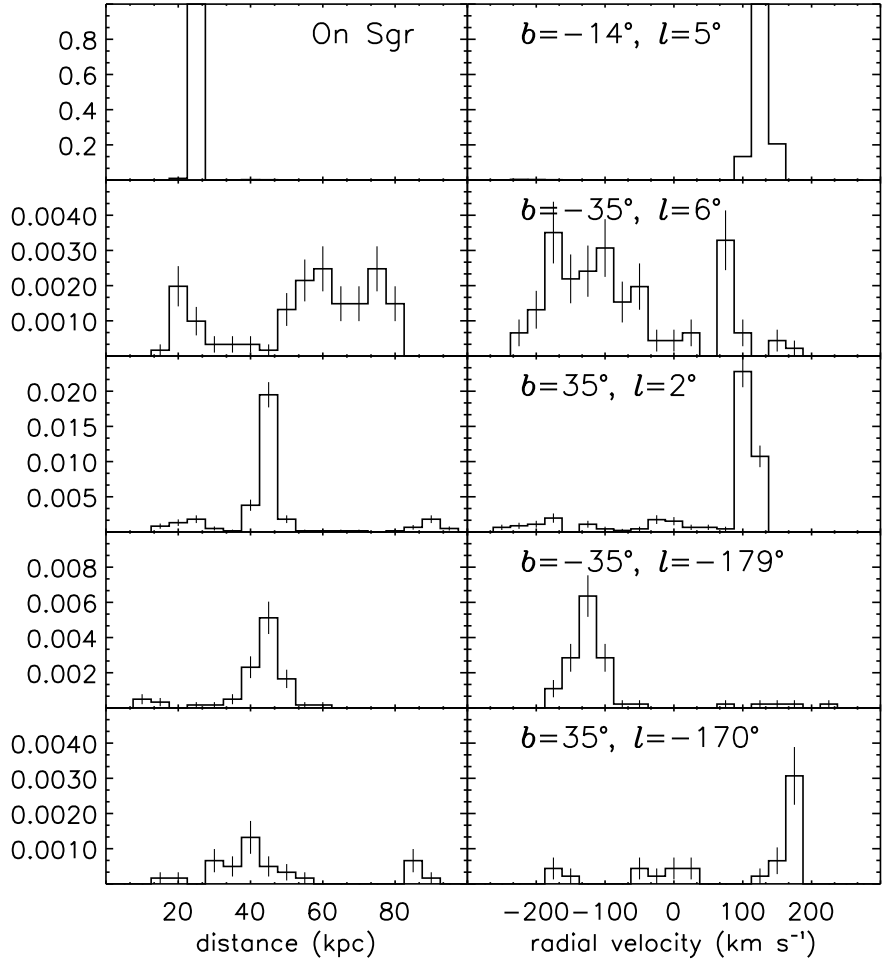,height=13.cm}
\end{center}
\caption{Number counts in $1 \times 1$ deg$^2$ normalized to the main
body of Sagittarius, which is shown in the top row. Distance bins are 5 kpc,
and radial velocity bins are 25$\kms$. All quantities are
heliocentric. Note that the debris reaches larger densities, and could
thus be more easily detectable, at $b \sim 35\ndeg$ for the stream in
the Galactic centre direction, and $b\sim-35\ndeg$ for the anticentre
stream.}
\label{fig:histograms}
\end{figure*}

\begin{figure*}
\begin{center}
\psfig{figure=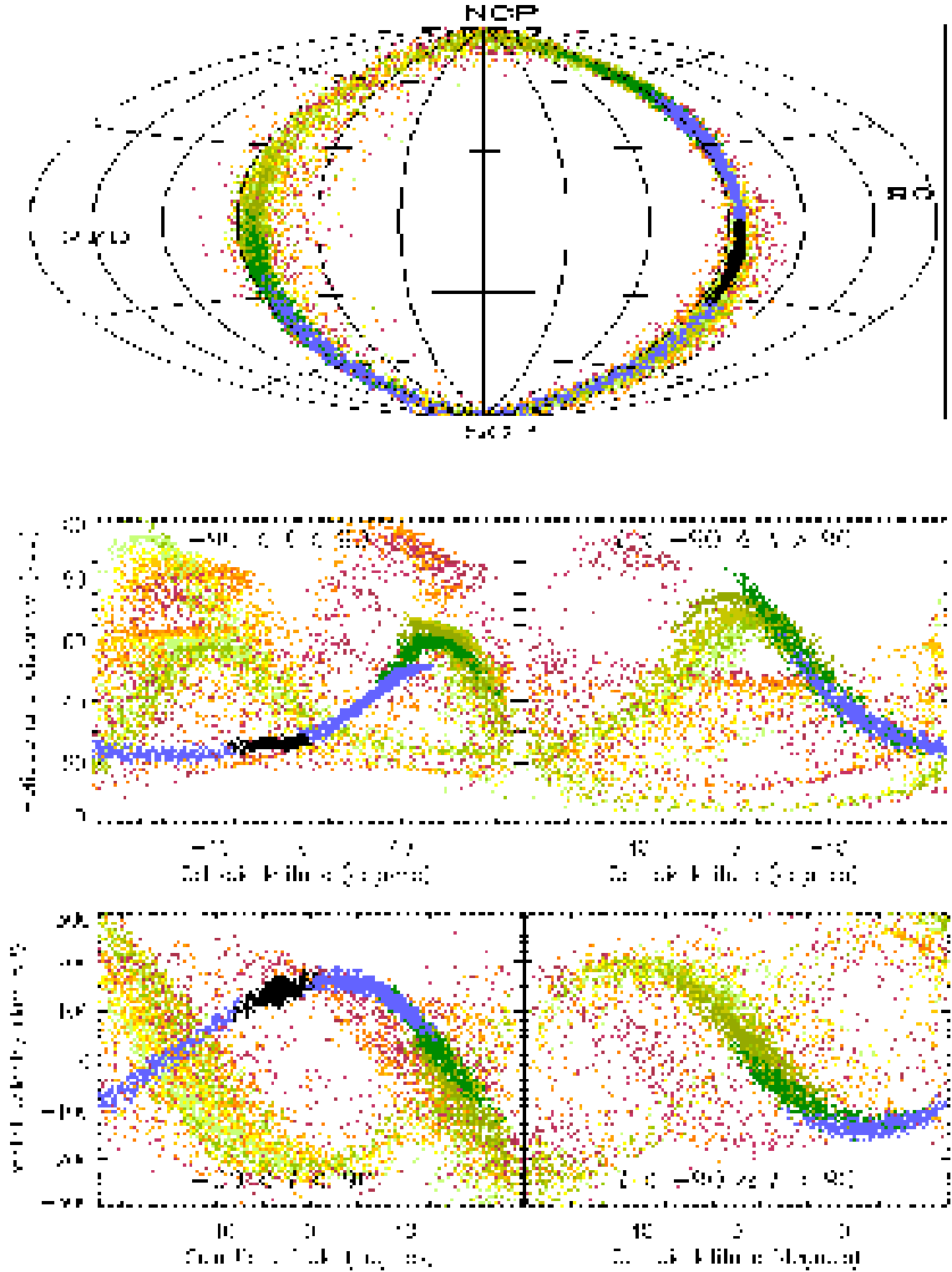,width=15.cm}
\caption{{\bf Top panel}: Distribution in the sky ($l,b$) of the
particles for our constant mass-to-light ratio model of Sagittarius after
12.5 Gyr. Different colours indicate material stripped off in
different passages. 
{\bf Central panel}: Heliocentric distance 
as a function of Galactic latitude, at the same time as the top panel, and
with the same colour coding. Note that
``streams'' formed early on are wider than the more recent ones. 
{\bf Bottom panel}: Heliocentric radial velocity as a function of Galactic
latitude, at the same time and using the same colour coding
as before.}
\label{fig:all_lat_long_dist_radvel_sgr}
\end{center}
\end{figure*}

In this section we concentrate for simplicity on Model I. We can use it to
predict star counts as a function of distance and radial velocity at
different points on the sky. We focus on fields along the path defined
by the orbit of Sgr, which is where we expect to find debris
streams. This is illustrated in Figure~\ref{fig:histograms}, where the
number counts are normalized to their values on the main body of our
simulated Sagittarius, as shown in the first row. We assume fields which are
$1\ndeg \times 1\ndeg$. For the distance, we use 5 kpc bins, whereas
for the radial velocity we take 25 $\kms$ bins.  Note that the
contrast of structures in the radial velocity counts are generally
larger than in the distance counts, indicating that it should be
easier to detect streams in velocity space rather than as density
inhomogeneities (see also Helmi \& White 1999). This is particularly
true considering the much greater relative precision of the velocity
measurements. Space density enhancements often occur near the orbital
turning points; several are seen as sharp features in the central
panel of Fig.~\ref{fig:all_lat_long_dist_radvel_sgr}.

Our model can also be used to predict where streams originating in
different mass loss events should be found.  This is illustrated in
Figure~\ref{fig:all_lat_long_dist_radvel_sgr} where different colours
indicate material lost at different pericentric passages. Note that
since the surface brightness of the unbound material decreases with
time, material lost in early passages is considerably more difficult
to detect than recent mass loss (for an axisymmetric potential the
time dependence is $1/t^2$, but if the potential may be considered as
nearly spherical the surface density will effectively decrease as
$1/t$; see Helmi \& White 1999).  The central panel (latitude
vs. heliocentric distance) explains why Sagittarius streams have been more
difficult to detect above the Galactic plane than below it, even
though the density contrast is higher for the northern streams (as
shown in the second and third panels of Fig.~\ref{fig:histograms}). 
From the left panel,$ -90\ndeg \le l \le 90\ndeg$, we see that the
stream of stars lost in the previous pericentric passage (shown in
blue) becomes more distant as we go north.  For example, at $b =
40\ndeg$, the stream is located approximately 50 kpc from the Sun. The
red giant clump visual magnitude at this distance would be roughly
19.3$^m$, compared to the 17.85$^m$ observed in the main body of Sagittarius.

\subsubsection{Comparison to data outside the main body of Sagittarius}

Even though we have constructed our models to reproduce the properties
of the main body of Sagittarius, it is nevertheless worthwhile to compare our
simulations to data sets which have claimed detections of Sagittarius debris.

\paragraph{Outer Structure of Sagittarius.}

\begin{figure}
\begin{center}
\psfig{figure=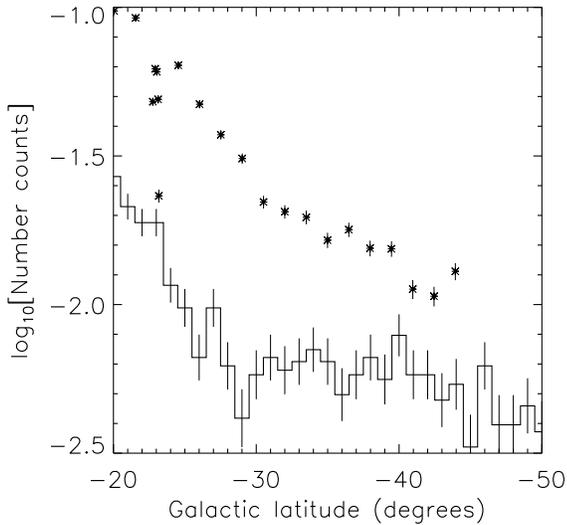,height=7.75cm}
\caption{Number counts along the major of axis of the remnant system,
outside its main body, within a strip of 3$\ndeg$ to 10$\ndeg$ in
longitude. The error bars indicate Poissonian noise in the number
counts. For comparison, we show the data by Mateo et al. (1998)
arbitrarily shifted.}
\label{fig:mateo98}
\end{center}
\end{figure}

Mateo et al. (1998) have traced Sagittarius material out to 30 degrees from
its nucleus: the globular cluster M54. They obtained deep photometric
data along the southeast extension of the major axis of Sagittarius.  In
Figure~\ref{fig:mateo98} we show the particle counts in our simulation
for the strip 3$\ndeg$ to 10$\ndeg$ in longitude, and spanning about
30$\ndeg$ in latitude outside the main remnant body. For comparison we
plot the data by Mateo and collaborators, shifted a few degrees in
latitude, and arbitrarily offset in number counts.  Thus qualitatively
we reproduce the break in the number counts profile. This change in
slope is indicative of the transition between material which is still
bound today and that lost in the last pericentric passage.

\paragraph{Star counts at $b=-40\ndeg$.}

Majewski et al. (1999) have claimed a detection of a possible stream
from Sagittarius at $b = -40\ndeg$ and $l = 11 \ndeg$, at a slightly smaller
heliocentric distance of $23$ kpc and with a radial velocity of the
order of $30\kms$. As they discuss, this velocity may be strongly
affected by contamination by other Galactic components. We note,
however, that we would predict a stream of stars (shown in blue) going
through this latitude and longitude with roughly the observed
distance, and with a radial velocity of $55 \kms$.  (See the central
and bottom left panels of Fig.~\ref{fig:all_lat_long_dist_radvel_sgr},
$ -90\ndeg \le l \le 90\ndeg$.).  As mentioned above, this stream is
formed mostly by material lost in the previous pericentric passage and
not three passages ago, as in the model of Johnston et al.  (1999).
This difference reflects the different orbital timescales in the two
models. The surface density of stars may be able to distinguish
between them; it is predicted to be higher in our case.

Unfortunately, Majewski and collaborators could not detect the
northern stream. They either did not reach the magnitude limit of
$19.3^m$ expected for the red giant clump, or were offset by a few
degrees from its expected location. Thus, for example, Majewski et
al. (1999) had a limiting magnitude of $\sim 21$ at $b = 41\ndeg$ and
$l = -6 \ndeg$, but $V \lesssim 19$ at $b = 41\ndeg$ and $l = 6
\ndeg$.  The actual stream in our model is predicted to go through $ l
\sim 1 \ndeg$ and to be about $2\ndeg$ wide.  Note that the width
prediction is more secure than the location since the motion of
Sagittarius in the $l$-direction is poorly constrained at present, 
although a flattened halo would make the streams wider.  
 
\paragraph{RR Lyrae found by the Sloan Digital Sky Survey.}

The Sloan Digital Sky Survey (SDSS) commissioning data has detected
148 candidate RR Lyrae stars in about 100 deg$^2$ of sky, along the
celestial equator ($-1.27\ndeg \!\le \delta \!\le 1.27\ndeg$), and
from $\alpha = 160.5\ndeg$ to $\alpha = 236.5\ndeg$ (Ivezic et al.,
2000).  Although the faint-magnitude limit of the SDSS would allow
them to detect RR Lyrae stars to large Galactocentric distances, they
find no candidates fainter than r*$\sim$ 20, i.e., farther than 65 kpc
from the Galactic center. The distribution of stars in their sample is
very inhomogeneous and shows a clump of over 50 stars at about 45 kpc
from the Galactic centre, which is also detected in the distribution
of nonvariable objects with RR Lyrae star colors.

\begin{figure*}
\begin{center}
\psfig{figure=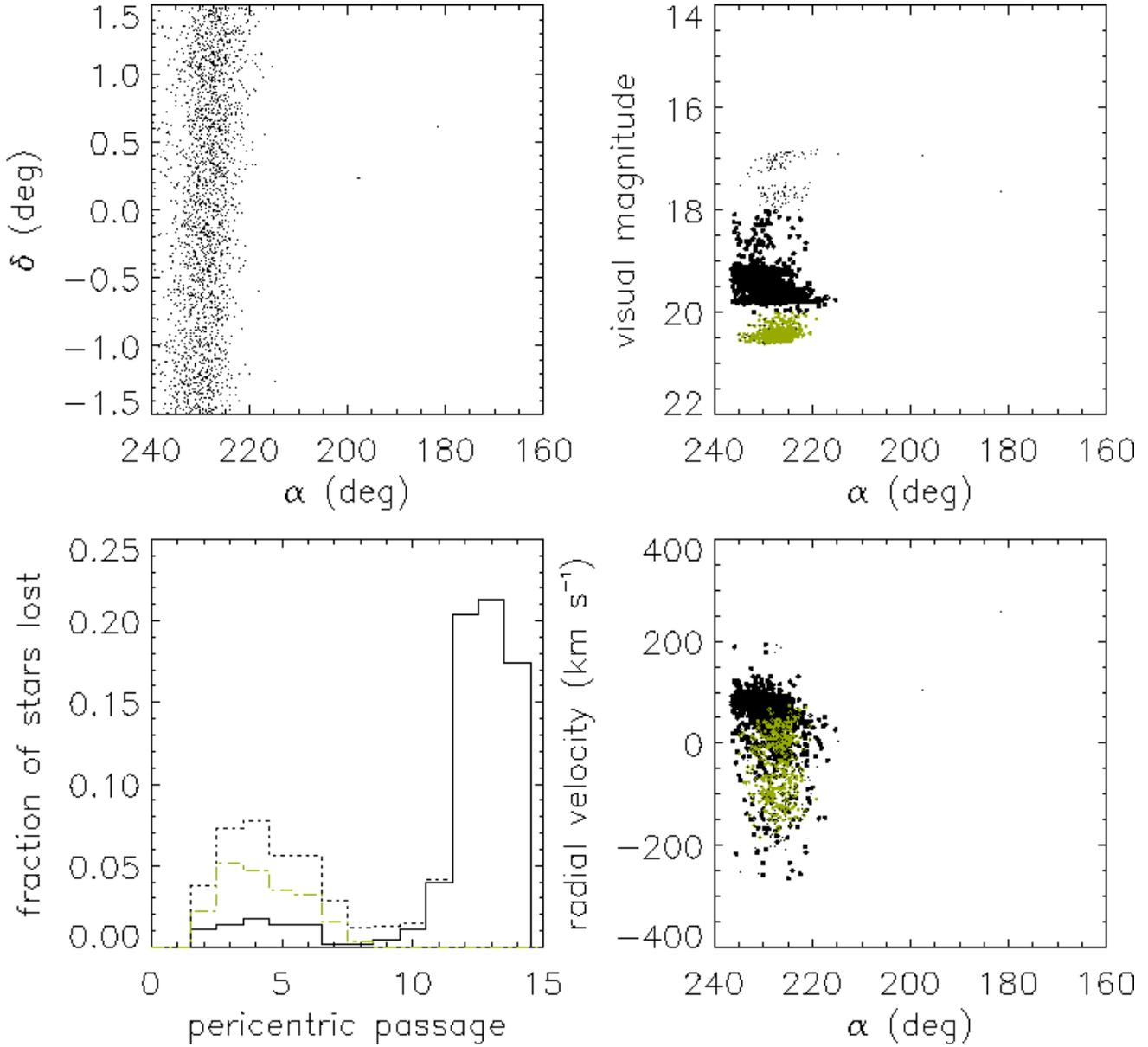,height=16cm}
\vspace*{1cm}\caption{The top left panel shows the region of the sky
analysed by the SDSS, where an excess of RR Lyrae has been
observed. The top right panel shows the distribution in apparent
magnitudes (i.e. distances for $M_V = 0.7^m$) of the particles in our
simulations falling in that region of the sky. We have colour coded
particles according to the range in distance: thick black dots
correspond to $18 \le V \le 20$, lighter black dots to $V \le 18$, and
grey diamonds to $V \ge 20$. Note that the first group is strongly
clustered around the magnitude range 19 -- 19.5, as found by the SDSS
for their RR Lyrae. The bottom left panel shows the distribution of
pericentric passages (i.e. times) when the particles became unbound
for each of the subgroups. The dotted histogram corresponds to all the
particles present in this field of the sky. We note here that there
are about twice as many particles which have been released in the last
3 Gyr, than earlier on. Most of the material in the first clump ($V
\sim 19 - 19.5$) became unbound in the 12th to 14th pericentric
passages, i.e. 1 -- 3 passages ago. On the contrary all particles in
the second clump ($V\sim 20.5$) became unbound in the first 7 passages.
Finally the bottom right panel shows the radial velocity distribution
with the same colour coding as before. We note that the stream appears
rather diffuse in velocity space and strongly clustered in space
because of the ``bunching up'' of the particles orbits that takes
place near their apocentres.}\label{fig:sloan}
\end{center}
\end{figure*}

By studying carefully Figure~\ref{fig:all_lat_long_dist_radvel_sgr},
and from our previous discussions, we are naturally led to believe
this substructure could be associated with the northern streams of
Sagittarius. In the upper left panel of Figure~\ref{fig:sloan} we see how, in
our simulations of Model I, a stream of material intersects the area
observed by SDSS.  The positions of the particles in our simulations are
in excellent agreement with those of the RR Lyrae candidates belonging
to the reported substructure. The upper right panel shows the
visual magnitude of the particles falling in the region of the sky
analysed by SDSS. We note here that there are basically two substructures
in this region: one at $V \sim 19.5^m$, and a second one, at a fainter
magnitude $V \sim 20.5^m$ (for $M_V = 0.7^m$ characteristic of RR Lyrae stars,
e.g. Layden et al. (1996)). The first lump clearly could correspond to
the substructure observed in the SDSS data. The material in
this lump is mostly formed by particles that were lost in recent
pericentric passages (i.e.  1 -- 3 Gyr ago) as shown in the bottom left
panel of Figure~\ref{fig:sloan}.

As Ivezic et al. (2000) discuss, they do not find any RR Lyrae stars
fainter than $V \sim 20^m$. This would be in apparent contradiction
with our results, (e.g. top right panel of
Fig.~\ref{fig:sloan}). However, we need to estimate how much material
we find in each lump, calibrate this number with respect to the number
of RR Lyrae in the lump observed by SDSS, and thereby determine how
many RR Lyrae SDSS could have missed. In the first lump we find 1264
particles, whereas the second has 362 particles. According to Ivezic
et al. (2000) the detection efficiency decreases rapidly between $V
\sim 20^m$, where it is fifty per cent, and $V \sim 21^m$ where it is
zero.  We here assume that for stars of $V \sim 20.5^m$ this
efficiency is about 15\%,  which means that only 54 of the 362
particles could, in principle, have been observed. Therefore, we
estimate that the ratio of unobserved to that of observed debris
material is 0.043 in this region of the sky.  Thus if SDSS found $\sim
50$ RR Lyrae belonging to the first substructure, it should have
detected $\sim 2.14 \pm 1.46 $ RR Lyrae in the fainter magnitude
range. This means that the failure to detect fainter RR Lyrae in this
region of the sky is barely significant in this context.  From this
perspective we cannot rule out that a second stream of debris material
is located at much larger distances (typically between 80 and 100 kpc
from the Sun, as shown in Figure~\ref{fig:all_lat_long_dist_radvel_sgr}).

Nevertheless the absence of a visible stream may be indicating that
this material could be dark-matter dominated. This second stream is
formed by particles that became unbound more than 7 Gyr ago. It
therefore corresponds to particles orbiting the outskirts of the
progenitor of Sagittarius. If this region of the system was dark-matter
dominated, such streams would remain unobservable.  Fainter data ($ V
\sim 20 - 21^m$) in this region of the sky could be crucial to constrain
the initial properties of the system, e.g. size, total
luminosity. This particular region of the sky should thus be explored
further!

\paragraph{Carbon stars by the APM}

The APM survey has detected about 75 high latitude carbon giants
presumably belonging to the halo. These stars being of intermediate
age, could trace streams that have recently become unbound from Sagittarius 
or from other Galactic satellites.  Ibata et al. (2000) have proposed
that a large fraction of the observed halo carbon stars belong to
Sagittarius tidal debris, since they preferentially occur near the great
circle of its orbit. Even though there are large uncertainties in the
determination of distances to these carbon stars, and the survey is
not complete, particularly in regions where we expect Sagittarius streams to
be present, this proposal clearly fits within the expectations for the
models we have developed here. 

\section{Conclusions}

We have found viable models for the Sagittarius dwarf galaxy with a wide
range of total luminosities and masses, and both with and without
extended dark halos. A purely stellar progenitor could be a King model
with a total velocity dispersion of 18.9~$\kms$, a core radius of
0.44~kpc and a tidal radius of 3~kpc. For the case where the
progenitor is embedded in an extended massive halo, the initial
stellar distribution follows a King profile with the same core radius,
a slightly larger total velocity dispersion of $23 \kms$ and similar
extent.  The dark-matter is more extended. The data available at
present only weakly constrain the total initial extent either of the
light or of the mass. The observed metallicity data, for example, are
consistent with an initial galaxy similar to either of our detailed
models, both of which would lie within the scatter of the
luminosity--size--velocity dispersion--metallicity distribution for
more distant dwarf spheroidal galaxies in the Local Group. Thus we see
no indication that Sagittarius is in any way anomalous.  Further work on the
debris streams of Sagittarius is needed to constrain better its initial total
luminosity, and to distinguish between purely stellar or dark-matter
dominated progenitors.

It is certainly encouraging that our models could reproduce the data
available both on the main body and on the debris streams. We wish to
stress however, that this does not mean that we have found the
``ultimate'' model. Other models with similar characteristics may also
exist. Alternatives would include progenitors with smaller stellar
masses or larger dark halos; flattened systems or with anisotropic
velocity distributions; or systems with a stellar disk and a spherical
dark halo (as proposed for the progenitors of dSph by Mayer et
al. (2000)). Moreover, our assumption of a rigid Galactic potential,
which does not vary in time over 12 Gyr, is clearly simplistic in view
of current models for the formation of structure in the Universe.
Only when we have a better estimate of the total luminosity of Sagittarius,
both in its main body, as well as on its streams, we will be able to
model it in greater detail. The present interest in the debris streams
of Sagittarius will help us understand not only the properties of what has
turned out to be just another dwarf spheroidal, but also the formation
history of our Galaxy. A complete map of the streams will, for
example, allow us to derive the Galactic potential (Johnston et
al. 1999). If these streams are less smooth or broader than expected,
this may indicate smaller scale structure present in the halo either now
or when this was assembled.

\vspace*{-0.25cm}
\section*{Acknowledgments}
We thank Pavel Kroupa and James Binney for comments on earlier
versions of this manuscript. We have enjoyed discussions with Heather
Morrison and Paul Harding. {\mbox CONICET}, Fundaci\'on Antorchas, 
DAAD--Fundaci\'on Antorchas, and EARA are acknowledged for financial
support. 

\label{lastpage}

\label{lastpage}

\end{document}